\documentclass[12pt]{article}
\usepackage{graphicx,amsmath,float}
\usepackage{units}
\usepackage{lineno}
\usepackage{xcolor}
\usepackage{afterpage}
\usepackage{multirow}
\usepackage{booktabs}

\newlength{\dinwidth}
\newlength{\dinmargin}
\def\lapproxeq{\lower .7ex\hbox{$\;\stackrel{\textstyle                                                    
<}{\sim}\;$}}                                                    
\def\gapproxeq{\lower .7ex\hbox{$\;\stackrel{\textstyle                                                    
>}{\sim}\;$}}                                                    
\def\be{\begin{equation}}                                                    
\def\ee{\end{equation}}                                                    
\def\bea{\begin{eqnarray}}                                                    
\def\eea{\end{eqnarray}}

\def\sh{\hat s}
\def\sh2{{\hat s}^2}

\begin{document}
                                                    
\titlepage                                                    
\begin{flushright}                              HIP-2021-33/TH \\                       
IPPP/21/45  \\  
LTH 1273 \\                             \vspace{0.3cm}                 
\today \\                                                    
\end{flushright} 
\vspace*{0.5cm}
\begin{center}                                                    
{\Large \bf Predictions of exclusive $\Upsilon$ photoproduction}\\
\vspace{0.5cm}
{\Large \bf at the LHC and future colliders }\\
\vspace*{1cm}
                                                   
C.~A.~Flett$^{a,b}$, S.~P.~Jones$^c$, A.~D.~Martin$^c$, M.~G.~Ryskin$^{c,d}$ and T.~Teubner$^e$\\                                                    
                                                   
\vspace*{0.5cm}  
\fontsize{10.47}{1}
$^a${\it Department of Physics, University of Jyv\"{a}skyl\"{a}, P.O. Box 35, 40014 University of Jyv\"{a}skyl\"{a}, Finland}\\
$^b${\it Helsinki Institute of Physics, P.O. Box 64, 00014 University of Helsinki, Finland}\\
$^c${\it Institute for Particle Physics Phenomenology, Durham University, Durham, DH1 3LE, U.K.} \\                              $^d${\it Petersburg Nuclear Physics Institute, NRC Kurchatov Institute, Gatchina, St.~Petersburg, 188300, Russia}  \\
$^e${\it Department of Mathematical Sciences, University of Liverpool, Liverpool, L69 3BX, U.K.}\\

\vspace*{1cm}                                                    
                                                    
\begin{abstract} 
\vspace*{0.2cm}  
The cross section for exclusive $\Upsilon$ ultraperipheral photoproduction at present and future colliders
is determined using the low $x$ gluon PDF extracted from an analysis of exclusive $J/\psi$ 
measurements performed at HERA and the LHC. Predictions are given at next-to-leading order in collinear factorisation over a wide $\gamma p$ centre-of-mass energy range, calculated assuming the non-relativistic approximation for the $\Upsilon$ wave function, and with skewing corrections incorporated.
\end{abstract}

\vspace*{0.5cm}                                                    
                                                    
\end{center}

The exclusive $\Upsilon$ photoproduction process, $\gamma p \rightarrow \Upsilon p$, was first measured in diffractive deep-inelastic-scattering (DIS) events by the ZEUS collaboration at the $ep$ HERA collider just short of 25 years ago~\cite{Breitweg:1998ki}. A subsequent measurement of this observable came from H1 at the start of the new millennium~\cite{Adloff:2000vm} and later again from ZEUS~\cite{Chekanov:2009zz}, extending the kinematic coverage of the datasets to larger values of the $\gamma p$ centre-of-mass energy. More recently, measurements of exclusive $\Upsilon$ production have been made by the LHCb collaboration in ultraperipheral $pp$ collisions at $pp$ centre-of-mass energies $\sqrt{s} = 7$ and $8$~TeV~\cite{Aaij:2015kea} and then, in the last few years, by the CMS collaboration in the $p\text{Pb}$ mode with centre-of-mass energy per nucleon pair of $\sqrt{s_{\text{NN}}} = 5.02$ TeV~\cite{Sirunyan:2018sav}. Forthcoming measurements at 8.16 TeV by CMS are anticipated~\cite{Naskar:2018frq, Dutta:2017izs}. 

In this short note we make predictions for the exclusive $\Upsilon$ photoproduction in a $\gamma p$ centre-of-mass energy range relevant for experiments past and present, and at the future Electron-Ion collider~(EIC) and the proposed LHeC and FCC. We use the collinear factorisation framework at NLO supplemented with a crucial $`Q_0$' subtraction~\cite{Jones:2016ldq}. We also employ the optimal factorisation scale $\mu_F = M_V/2$ which reduces the scale dependence of the result thanks to the resummation of double logarithmic, $(\alpha_s \ln(1/x) \ln(\mu_F))^n$, terms. Here, $M_V$ is the mass of the vector meson.
For very low $x$, we use the gluon PDF determined by a fit to exclusive $J/\psi$ photoproduction data in~\cite{Flett:2020duk}, which does not include any $\Upsilon$ data. 

\begin{figure} [t]
\begin{center}
\includegraphics[width=0.4\textwidth]{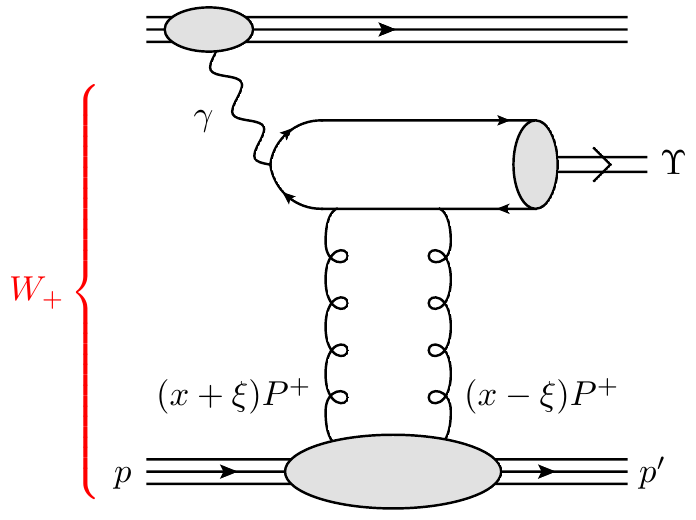}
\qquad
\includegraphics[width=0.4\textwidth]{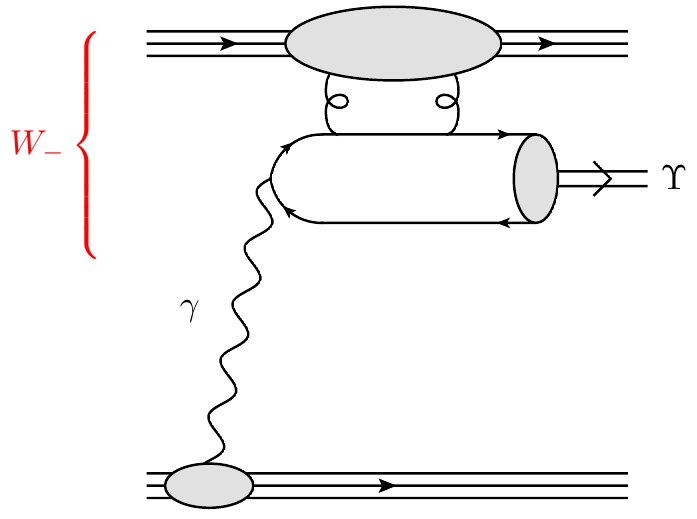}
\caption{\sf{The two diagrams describing exclusive $\Upsilon$ production, $pp \rightarrow p \Upsilon p$, at the LHC. The $W_+$ and $W_-$ contributions arise in the ultraperipheral description of the $\gamma p \rightarrow \Upsilon p$ subprocess, see the text for details. In the $p$Pb mode, either the upper or lower proton is replaced by a Pb-ion.
  }}
\label{fig:f2}
\end{center}
\end{figure}

Let us briefly recall our formalism. We work at NLO within 
the collinear factorisation scheme and express the amplitude for exclusive $\Upsilon$ photoproduction as
\begin{equation}
    A \propto \langle O_1 \rangle_{\Upsilon} \int_{-1}^1 \text{d}x \left( C_g(x,\xi) F_g(x,\xi) + \sum_{q=u,d,s} C_q(x,\xi) F_q(x,\xi) \right),
    \label{amp}
\end{equation}
where $F_g$ and $F_q$ are Generalised Parton Distributions~(GPDs), $C_g$ and $C_q$ are coefficient functions and $x-\xi,~x+\xi$ are parton momentum fractions in the lightcone direction $P^+$. The dependence on the factorisation scale $\mu_F$ and on the four-momentum transfer squared, $t$, is not shown. The set-up is shown in Fig.~\ref{fig:f2}. The non-relativistic QCD~(NRQCD) matrix element $\langle O_1 \rangle_{\Upsilon} $ is fixed by the experimental value of the $\Upsilon \rightarrow \mu^+ \mu^-$ decay width.
In~\cite{Hoodbhoy:1996zg}, it was demonstrated that relativistic corrections to the $J/\psi$ wave function suppress the cross section by $\sim 6\%$. For $\Upsilon$ production, due to the larger quark mass, this suppression is expected to be a smaller effect. 

It was argued in~\cite{Jones:2015nna} (in the context of exclusive $J/\psi$ production but nonetheless generally for the exclusive production of heavy vector mesons) that the factorisation scale choice $\mu_F = M_{V}/2$ resums the logarithmically enhanced terms $\sim (\alpha_s \ln(1/x) \ln( \mu_F ) )^n$ in the NLO amplitude at small $x$. Moreover, in~\cite{Jones:2016ldq} the double counting of contributions $\mathcal O(Q_0^2/M_V^2) \sim \mathcal O(1)$ in the low parton transverse momentum $k_t < Q_0$ domain was eliminated, where $Q_0$ is the PDF input scale. Together, these effects resulted in an NLO correction smaller than the Born contribution, and a reduced dependence on the factorisation scale. 

The Shuvaev integral transform~\cite{Shuvaev:1999fm, Shuvaev:1999ce, Martin:2009zzb} is used to relate the conventional collinear PDFs to the GPDs at small $x$. This provides sufficient accuracy $\sim \mathcal O(x)$ at NLO in the low $x$ domain. As the transform is not valid in the timelike sub-region $|x|<\xi$ of the integration domain, we use eq.~\eqref{amp} to extract the {\it imaginary} part of the amplitude only. In this region, the imaginary part of the coefficient functions are zero. The real part is restored at the level of the total amplitude via a dispersion relation which in the high energy limit (for an even signature amplitude) can be written in the simplified form,
\be
\rho ~~=~~ \frac{{\rm Re}A}{ {\rm Im}A}~~=~~{\rm tan}\left(\frac{\pi}{2}~\frac{\partial \ln{\rm Im}A/W^2}{\partial \ln W^2}\right),
\ee
see e.g.~\cite{Ryskin:1995hz}. Here, $W$ is the $\gamma p$ centre-of-mass energy.
The cross section, differential in $t$, evaluated at zero momentum transfer in the $t$-channel, is given by
\begin{equation}
 \frac{\text{d} \sigma}{\text{d} t} \left(\gamma p \rightarrow \Upsilon p\right)\biggl|_{t=0} = \frac{(\text{Im} A)^2 (1 + \rho^2)}{16 \pi W^4 }.
\end{equation}
To describe data integrated over $t$ we assume that the  cross  section  depends  exponentially  on $t$, that is  $\sigma \sim \exp(-B |t|)$.   The  energy-dependent slope parameter, $B$, is given by the Regge motivated parametrisation
\begin{equation}
B(W) = \left (B_0 + 4 \alpha'_P \ln \left(\frac{W}{W_0} \right) \right)\,\,\,\text{GeV}^{-2},
\end{equation}
where $B_0 = 4.63\, \text{GeV}^{-2}$ is the relevant intercept for $\Upsilon$ production, and the pomeron slope $\alpha'_P = 0.06 \, \text{GeV}^{-2}$ and $W_0 = 90\, \text{GeV}$. This parametrisation grows more slowly with $W$ than that in~\cite{Aaij:2015kea} and is based on Model 4 of~\cite{Khoze:2013dha}, which fits a wider variety of elastic $pp$ scattering data.

\afterpage{
\begin{figure} [t]
\begin{center}
\includegraphics[width=0.8\textwidth]{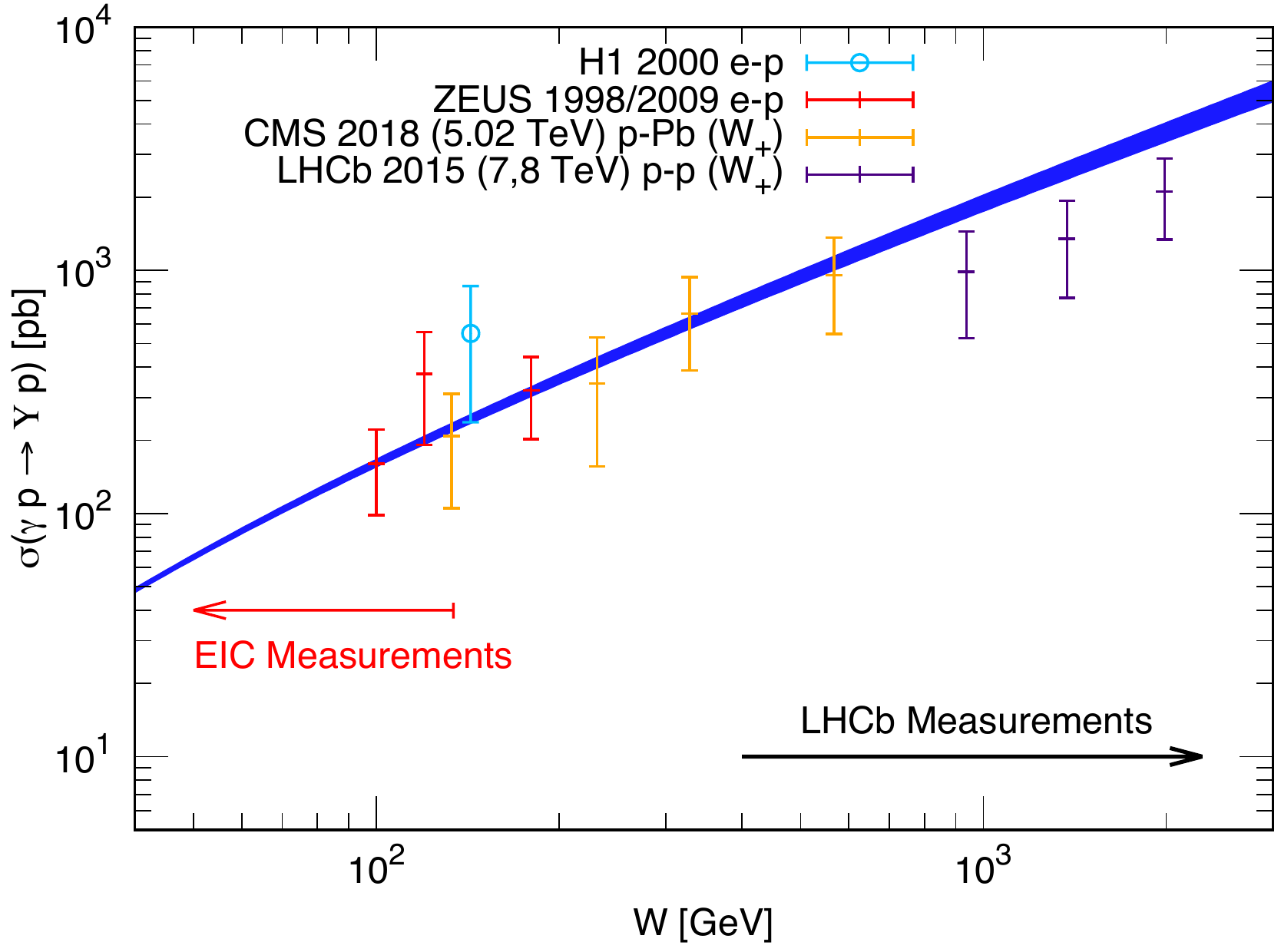}
\caption[..]{\sf{The cross section prediction for the $\gamma p \rightarrow \Upsilon p$ process based on a DGLAP evolved gluon PDF obtained from a fit to exclusive $J/\psi$ data \cite{Flett:2020duk}, and quark PDFs from global analyses, here NNPDF3.0\footnotemark[1]. We stress that no fit to the available $\Upsilon$ data shown is made at this stage and that the width of the blue band represents the propagation of the $\pm 1 \sigma$ uncertainties of the exclusive $J/\psi$ fit parameters. Shown for comparison purposes are the currently available exclusive $\Upsilon$ data~\cite{Breitweg:1998ki, Adloff:2000vm, Chekanov:2009zz, Aaij:2015kea, Sirunyan:2018sav} as well as the projected kinematic coverage of this observable for the future EIC in its highest energy configuration. (The apparent discrepancy of the LHCb data with our predictions may be explained as discussed in the penultimate paragraph of the paper.)
  }}
\label{fig:cross}
\end{center}
\end{figure}
\footnotetext[1]{The shape and normalisation of the gluon PDF at the matching point $x=10^{-3}$ and the $\Upsilon$ scale from recent global PDF analyses are similar. 
Our prediction therefore does not strongly depend on the PDF set used for the matching. 
}
}

Figure~\ref{fig:cross} displays our predictions for the $\gamma p \rightarrow \Upsilon p$ cross section at NLO. It is based on using, as input, the behaviour of the low $x$ gluon PDF determined from an analysis of exclusive $J/\psi$ data,
with the quark PDFs taken from the global analyses. Note that at NLO, the evolution of the gluon and singlet quark PDFs, relevant here, are coupled.
However, it was shown in~\cite{Flett:2019pux} that after the $Q_0$ subtraction, the quark contribution to eq.~\eqref{amp} is negligible. Therefore, our separate treatment of the quark and gluon PDFs is justified at this level of accuracy. Indeed, within the DGLAP approach (with strong $k_t$ ordering), the $k_t$ of the light quarks is smaller than $\mu_F$ (since the quark contribution is separated from the outgoing $\Upsilon$ meson by at least one step of DGLAP evolution - only the gluons may participate in the hard scattering event). This means that practically the whole quark contribution comes from the region $k_t < Q_0$ and therefore, after the $Q_0$ subtraction, is more or less absorbed into
the input PDF. We have checked in our approach that inclusion of the quarks leads to about a $1\%$ enhancement of the cross section in the low $x$ domain and so our prediction is driven by the gluon distribution.  Explicitly, in~\cite{Flett:2020duk} it was found that fitting a power ansatz for the low $x$ gluon PDF, $xg \sim x^{-\lambda}$, to the exclusive $J/\psi$ data from LHCb at 7 and 13 TeV~\cite{Aaij:2014iea, Aaij:2018arx}, and to the HERA data with $x \lapproxeq 0.001$~\cite{Chekanov:2002xi, Chekanov:2004mw, Aktas:2005xu, Alexa:2013xxa}, gave an excellent description with a $\chi_{\text{min}}^2/\text{d.o.f} \approx 1$. Moreover, the gluon PDF inspired by the double-logarithmic approximation~(DLA) 
\begin{equation}
xg^{}(x,\mu_0^2)~\sim~
(1-x)~x^{-a}\left(\frac{\mu_0^2}{q^2_0}\right)^{-0.2}~{\rm
exp}\left[\sqrt{16(N_c/\beta_0)\ln (1/x)\ln G}\right]
\label{DL1}
\ee
\be
{\rm with}~~~~G~=~\frac{\ln (\mu_0^2/\Lambda^2_{\rm QCD})}{\ln (q_0^2/\Lambda^2_{\rm
QCD})}~
\label{DL2}
\end{equation}
was also used and gave a similar fit quality. Here, $\Lambda_{\text{QCD}}=200\,\text{MeV}$ and $q_0^2 = 1\,\text{GeV}^2$, with $\beta_0 = 9$ for three light quark flavours. In the low $x$ region, the expected $x$ dependence of the gluon density follows a pure power law, but evolution in the scale quickly modifies this behaviour, resulting in a steeper gluon at larger $\mu_0^2$. The exponential term in~\eqref{DL1} resums the double logarithmic terms $\sim (\alpha_s \ln (1/x) \ln(\mu_F) )^n$ to all orders in $n$ and so we find that, to good accuracy, the NLO DGLAP low $x$ evolution in the $Q^2$ interval from 2 to about 30 $\text{GeV}^2$ is reproduced. The DLA parametrisation therefore mimics DGLAP evolution in a region that includes the $\Upsilon$ photoproduction scale $\mu_0^2 = (M_V/2)^2 \simeq 22.4\,\text{GeV}^2$.

We use~\eqref{DL1} and~\eqref{DL2} at the $\Upsilon$ scale, taking the slope and normalisation of $xg$ from the DLA fit made to the exclusive $J/\psi$ data.  To obtain the effective power of $\lambda$ we fit the resulting
grid of values over the range of $x$ corresponding to the $W$ range in Fig.~\ref{fig:cross}. The normalisation is fixed by matching onto the global partons at $x = 10^{-3}$. In this way we obtain $\lambda \approx 0.24$. We have checked that this is in line with the effective power growth of the gluon density from the NNPDF3.0 global parton set~\cite{NNPDF:2014otw} at $\mu_0^2 \approx 22.4\,\text{GeV}^2$ in the $x$ range considered. Moreover, we have verified that \texttt{APFEL++}~\cite{Bertone:2017gds} gives the same power behaviour when we DGLAP evolve our low $x$ power ansatz fitted to the exclusive $J/\psi$ data, $xg \sim x^{-\lambda}$ with $\lambda \approx 0.14$, from the $J/\psi$ scale to the $\Upsilon$ scale. 

We emphasise that the prediction shown in Fig.~\ref{fig:cross} is based only on DGLAP evolving a previously obtained gluon distribution. The data are shown just for comparison purposes and are not included in any fit at this stage. The width of the shaded band gives the $1 \sigma$ uncertainty from the $J/\psi$ experimental data used in the gluon PDF fits but does not account for theoretical uncertainties. 

The data for exclusive $\Upsilon$ production via ultraperipheral $pp$ and $p$Pb collisions from LHCb and CMS respectively can be used to estimate exclusive $\Upsilon$ photoproduction, $\gamma p \rightarrow \Upsilon p$, via an unfolding procedure as described in~\cite{Aaij:2015kea}. Broadly speaking, the cross section data from LHCb (CMS) measured differentially in bins of rapidity for $pp \rightarrow p \Upsilon p$ ($p$Pb $\rightarrow p\Upsilon$Pb) collisions can be used to find a cross section for the underlying $\gamma p \rightarrow \Upsilon p$ subprocess.  

In the absence of forward proton tagging in $pp$ collisions at the LHCb,  there is the ambiguity regarding which proton acted as the photon emitter and which as the target so, for a given rapidity $Y$, there are two different $\gamma p$ subprocesses with different centre-of-mass energies $W_{\pm}^2 = M_{\Upsilon}\,\sqrt{s}\,\exp(\pm |Y|)$ that contribute, as illustrated in Fig.~1. The interference effect between the two subprocesses is small and will be neglected in the following. To be specific, exclusive $\Upsilon$ production in ultraperipheral $pp$ collisions, $\text{d} \sigma(pp)/\text{d}Y$, can therefore be expressed in terms of the exclusive photoproduction cross sections $\sigma_{\pm}( \gamma p)$, for the subprocess $\gamma p \rightarrow \Upsilon p$ at the two energies $W_{\pm}$, by the equation

\begin{equation}
\frac{\text{d} \sigma (pp)}{\text{d} Y} = S^2(W_+)   \left(k_+  \frac{\text{d} n}{\text{d} k}_+ \right) \sigma_+(\gamma p) + S^2(W_-) \left( k_-  \frac{\text{d} n}{\text{d} k}_- \right) \sigma_-(\gamma p),
\label{dsigma}
\end{equation}
where $k_{\pm} \text{d}n/\text{d}k_{\pm}$ are photon fluxes and $S^2(W_{\pm})$ are survival factor corrections, accounting for the probability that the rapidity gap is not populated by additional soft interactions involving the initial state proton.  

 While both $W_+$ and $W_-$ contributions exist in the $p$Pb configuration too, experimentally the ambiguity of the photon emitter can be somewhat alleviated by detection of neutrons from the Pb-ion using zero degree calorimeters, as employed by CMS. In the experimental analyses, the $W_-$ component is treated as a systematic uncertainty~\cite{Aaij:2015kea} or as a background~\cite{Sirunyan:2018sav}.

Below, we compare the choice of photon flux and survival factor combination taken from~\cite{Jones:2013pga} with that constructed using the more accurate photon flux from~\cite{Budnev:1975poe}. Survival factors compatible with the photon flux presented in~\cite{Budnev:1975poe} are given in Table~\ref{survive}. 

\begin{table} [htbp]
\centering
\begin{tabular}{ccccccc}
\toprule
\multicolumn{1}{c}{$Y$} &
\multicolumn{2}{c}{7 TeV}    &
\multicolumn{2}{c}{8 TeV} &
\multicolumn{2}{c}{13 TeV} \\
\cmidrule(lr){2-3}
\cmidrule(lr){4-5}
\cmidrule(lr){6-7}

&
\multicolumn{1}{c}{$S^2(W_+)$} &
\multicolumn{1}{c}{$S^2(W_-)$}     &
\multicolumn{1}{c}{$S^2(W_+)$} &
\multicolumn{1}{c}{$S^2(W_-)$}  &
\multicolumn{1}{c}{$S^2(W_+)$}  &
\multicolumn{1}{c}{$S^2(W_-)$}  \\
\midrule
0.125 & 0.806 & 0.815 & 0.809 & 0.817 & 0.818 & 0.826 \\
0.375 & 0.796 & 0.823 & 0.799 & 0.825 & 0.810 & 0.833\\
0.625 & 0.785 & 0.830 & 0.789 & 0.832 & 0.801 & 0.839\\
0.875 & 0.773 & 0.837 & 0.777 & 0.839 & 0.791 & 0.845\\
1.125 & 0.760 & 0.843 & 0.765 & 0.845 & 0.781 & 0.850\\
1.375 & 0.745 & 0.849 & 0.751 & 0.851 & 0.769 & 0.855\\
1.625 & 0.728 & 0.854 & 0.735 & 0.856 & 0.756 & 0.860\\
1.875 & 0.709 & 0.860 & 0.717 & 0.861 & 0.741 & 0.865\\
2.125 & 0.688 & 0.864 & 0.697 & 0.865 & 0.724 & 0.869\\
2.375 & 0.664 & 0.869 & 0.674 & 0.870 & 0.706 & 0.873\\
2.625 & 0.637 & 0.873 & 0.648 & 0.874 & 0.684 & 0.877\\  
2.875 & 0.606 & 0.877 & 0.619 & 0.877 & 0.661 & 0.880\\
3.125 & 0.571 & 0.880 & 0.586 & 0.881 & 0.634 & 0.883\\
3.375 & 0.532 & 0.884 & 0.549 & 0.884 & 0.604 & 0.886\\
3.625 & 0.488 & 0.887 & 0.507 & 0.887 & 0.569 & 0.889\\
3.875 & 0.441 & 0.890 & 0.462 & 0.890 & 0.531 & 0.892\\
4.125 & 0.392 & 0.893 & 0.413 & 0.893 & 0.488 & 0.895\\
4.375 & 0.341 & 0.896 & 0.363 & 0.896 & 0.441 & 0.897\\
4.625 & 0.290 & 0.898 & 0.312 & 0.899 & 0.392 & 0.900\\
4.875 & 0.243 & 0.901 & 0.262 & 0.901 & 0.340 & 0.902\\
5.125 & 0.200 & 0.903 & 0.217 & 0.903 & 0.289 & 0.904\\
5.375 & 0.164 & 0.905 & 0.177 & 0.906 & 0.240 & 0.906\\
5.625 & 0.133 & 0.907 & 0.144 & 0.908 & 0.196 & 0.908\\
5.875 & 0.109 & 0.910 & 0.117 & 0.910 & 0.158 & 0.910\\
\bottomrule
\end{tabular}
\caption{\sf{Rapidity gap survival factors $S^2$ for exclusive $\Upsilon$ production, $pp \rightarrow p + \Upsilon + p$, as a function of the $\Upsilon$ rapidity $Y$ for $pp$ centre-of-mass energies of 7~TeV, 8~TeV and 13~TeV. The columns labelled $S^2(W_{\pm})$ give the gap survival factors for the two independent $\gamma p \rightarrow \Upsilon p$ subprocesses at different $\gamma p$ centre of mass energies $W_{\pm}$. }}
\label{survive}
\end{table}
For $\text{d} \sigma (pp)/\text{d} Y $, we observe that this choice produces a difference of at most 5\% at the maximum forward rapidity $Y \sim 4.5$ for exclusive $J/\psi$ production at LHCb. At an even larger rapidity $Y \sim 5$ (beyond the acceptance of LHCb), this difference increases to $\sim 25\%$. 
The mass of the $\Upsilon$ is $\sim 3$ times that of the $J/\psi$ and so (with $k_+ \propto M_V$ and $W_+ \propto \sqrt{M_V})$, the typical photon energy in exclusive $\Upsilon$ production is now much larger than in exclusive $J/\psi$ production and we enter the region where the approximation of the photon flux presented in~\cite{Jones:2013pga} breaks down at much lower rapidities (within the acceptance of LHCb and CMS). The large $W_+$ data points from LHCb shown in Fig.~\ref{fig:cross} (where the photon flux and $S^2$ from~\cite{Jones:2013pga} were used) are shifted towards our prediction if the photon flux and survival factor combination constructed based on the work presented in~\cite{Budnev:1975poe} is used in~\eqref{dsigma}.  To emphasise, though the photon flux used in~\cite{Jones:2013pga} is adequate for exclusive $J/\psi$ production in $pp$ collisions for $Y<4.5$, for higher $Y$ and particularly for exclusive $\Upsilon$ production we should use the more accurate photon flux of~\cite{Budnev:1975poe}.

In summary, using the framework built and developed in~\cite{Jones:2016ldq, Flett:2019pux, Flett:2020duk}, we have predicted the cross section for exclusive $\Upsilon$ production at HERA and in ultraperipheral collisions at the LHC, using a low $x$ gluon parametrisation
extracted from HERA and LHC exclusive $J/\psi$ production data. More precise exclusive $\Upsilon$ data are anticipated from LHCb, with their HERSCHEL detector now employed, in $pp$ collisions, and from CMS in $p$Pb collisions, as well as in the upcoming High-Luminosity phase of the LHC and the $ep$ programme of the EIC. While the statistics achievable for $\Upsilon$ production may be more limited than that for $J/\psi$, the theoretical uncertainties are under better control. A combined fit to $\Upsilon$ together with the $J/\psi$ and $\psi(2S)$ data would therefore be desirable in the future. All such data will increase our understanding of the underlying theoretical mechanisms at play in these interactions and, importantly, lead to an improved understanding of the behaviour of the gluon distribution at small $x$.
This programme will also require a more complete theoretical treatment of exclusive $\Upsilon$ production in $p$Pb and Pb$p$ collisions, accounting for the possible proton rescattering inside the Pb-ion, which we leave for future work.

\section*{Acknowledgements}

We would like to thank Dipanwita Dutta and Kousik Naskar for encouraging us to make these predictions. CAF is supported by the Helsinki Institute of Physics core funding project QCD-THEORY (project 7915122). SPJ is supported by a Royal Society University Research Fellowship (Grant URF/R1/201268). The work of TT is supported by the STFC Consolidated Grant ST/T000988/1.

\bibliographystyle{unsrt}
\bibliography{references.bib}

\begin{thebibliography}{10}

\bibitem{Breitweg:1998ki}
J.~Breitweg et~al.
\newblock {Measurement of elastic $\Upsilon$ photoproduction at HERA}.
\newblock {\em Phys. Lett. B}, 437:432--444, 1998.

\bibitem{Adloff:2000vm}
C.~Adloff et~al.
\newblock {Elastic photoproduction of $J /\psi$ and $\Upsilon$ mesons at HERA}.
\newblock {\em Phys. Lett. B}, 483:23--35, 2000.

\bibitem{Chekanov:2009zz}
S.~Chekanov et~al.
\newblock {Exclusive photoproduction of $\Upsilon$ mesons at HERA}.
\newblock {\em Phys. Lett. B}, 680:4--12, 2009.

\bibitem{Aaij:2015kea}
Roel Aaij et~al.
\newblock {Measurement of the exclusive \ensuremath{\Upsilon} production
  cross-section in pp collisions at $ \sqrt{s}=7 $ TeV and 8 TeV}.
\newblock {\em JHEP}, 09:084, 2015.

\bibitem{Sirunyan:2018sav}
Albert~M Sirunyan et~al.
\newblock {Measurement of exclusive $\Upsilon$ photoproduction from protons in
  pPb collisions at $\sqrt{s_\mathrm{NN}} =$ 5.02 TeV}.
\newblock {\em Eur. Phys. J. C}, 79(3):277, 2019.

\bibitem{Naskar:2018frq}
Kousik Naskar, Dipanwita Dutta, and Pradeep Sarin.
\newblock {Background study of $\Upsilon$ photoproduction in pPb collisions at
  8.16 TeV with CMS experiment}.
\newblock {\em DAE Symp. Nucl. Phys.}, 63:958--959, 2018.

\bibitem{Dutta:2017izs}
Dipanwita Dutta, Kousik Naskar, Ruchi Chudasama, and Pradeep Sarin.
\newblock {Study of $\Upsilon$ photoproduction in pPb collisions at 8.16 TeV
  with CMS experiment}.
\newblock {\em DAE Symp. Nucl. Phys.}, 62:964--965, 2017.

\bibitem{Jones:2016ldq}
S.~P. Jones, A.~D. Martin, M.~G. Ryskin, and T.~Teubner.
\newblock {The exclusive $J/\psi$ process at the LHC tamed to probe the low $x$
  gluon}.
\newblock {\em Eur. Phys. J. C}, 76(11):633, 2016.

\bibitem{Flett:2020duk}
C.~A. Flett, A.~D. Martin, M.~G. Ryskin, and T.~Teubner.
\newblock {Very low $x$ gluon density determined by LHCb exclusive $J/\psi$
  data}.
\newblock {\em Phys. Rev. D}, 102:114021, 2020.

\bibitem{Hoodbhoy:1996zg}
Pervez Hoodbhoy.
\newblock {Wave function corrections and off forward gluon distributions in
  diffractive $J / \psi$ electroproduction}.
\newblock {\em Phys. Rev. D}, 56:388--393, 1997.

\bibitem{Jones:2015nna}
S.~P. Jones, A.~D. Martin, M.~G. Ryskin, and T.~Teubner.
\newblock {Exclusive $J/\psi$ and $\Upsilon$ photoproduction and the low $x$
  gluon}.
\newblock {\em J. Phys. G}, 43(3):035002, 2016.

\bibitem{Shuvaev:1999fm}
A.~Shuvaev.
\newblock {Solution of the off forward leading logarithmic evolution equation
  based on the Gegenbauer moments inversion}.
\newblock {\em Phys. Rev. D}, 60:116005, 1999.

\bibitem{Shuvaev:1999ce}
A.~G. Shuvaev, K.~J. Golec-Biernat, A.~D. Martin, and M.~G. Ryskin.
\newblock {Off diagonal distributions fixed by diagonal partons at small $x$
  and $\xi$}.
\newblock {\em Phys. Rev. D}, 60:014015, 1999.

\bibitem{Martin:2009zzb}
A.~D. Martin, C.~Nockles, M.~G. Ryskin, A.~G. Shuvaev, and T.~Teubner.
\newblock {Generalised parton distributions at small $x$}.
\newblock {\em Eur. Phys. J. C}, 63:57--67, 2009.

\bibitem{Ryskin:1995hz}
M.~G. Ryskin, R.~G. Roberts, Alan~D. Martin, and E.~M. Levin.
\newblock {Diffractive $J / \psi$ photoproduction as a probe of the gluon
  density}.
\newblock {\em Z. Phys. C}, 76:231--239, 1997.

\bibitem{Khoze:2013dha}
V.~A. Khoze, A.~D. Martin, and M.~G. Ryskin.
\newblock {Diffraction at the LHC}.
\newblock {\em Eur. Phys. J. C}, 73:2503, 2013.

\bibitem{Flett:2019pux}
C.~A. Flett, S.~P. Jones, A.~D. Martin, M.~G. Ryskin, and T.~Teubner.
\newblock {How to include exclusive $J/\psi$ production data in global PDF
  analyses}.
\newblock {\em Phys. Rev. D}, 101(9):094011, 2020.

\bibitem{Aaij:2014iea}
Roel Aaij et~al.
\newblock {Updated measurements of exclusive $J/\psi$ and $\psi$(2S) production
  cross-sections in pp collisions at $\sqrt{s}=7$ TeV}.
\newblock {\em J. Phys. G}, 41:055002, 2014.

\bibitem{Aaij:2018arx}
Roel Aaij et~al.
\newblock {Central exclusive production of $J/\psi$ and $\psi(2S)$ mesons in
  $pp$ collisions at $\sqrt{s}=13~$TeV}.
\newblock {\em JHEP}, 10:167, 2018.

\bibitem{Chekanov:2002xi}
S.~Chekanov et~al.
\newblock {Exclusive photoproduction of $J / \psi$ mesons at HERA}.
\newblock {\em Eur. Phys. J. C}, 24:345--360, 2002.

\bibitem{Chekanov:2004mw}
S.~Chekanov et~al.
\newblock {Exclusive electroproduction of $J/\psi$ mesons at HERA}.
\newblock {\em Nucl. Phys. B}, 695:3--37, 2004.

\bibitem{Aktas:2005xu}
A.~Aktas et~al.
\newblock {Elastic $J/\psi$ production at HERA}.
\newblock {\em Eur. Phys. J. C}, 46:585--603, 2006.

\bibitem{Alexa:2013xxa}
C.~Alexa et~al.
\newblock {Elastic and Proton-Dissociative Photoproduction of $J/\psi$ Mesons
  at HERA}.
\newblock {\em Eur. Phys. J. C}, 73(6):2466, 2013.

\bibitem{NNPDF:2014otw}
Richard~D. Ball et~al.
\newblock {Parton distributions for the LHC Run II}.
\newblock {\em JHEP}, 04:040, 2015.

\bibitem{Bertone:2017gds}
Valerio Bertone.
\newblock {APFEL++: A new PDF evolution library in C++}.
\newblock {\em PoS}, DIS2017:201, 2018.

\bibitem{Jones:2013pga}
S.~P. Jones, A.~D. Martin, M.~G. Ryskin, and T.~Teubner.
\newblock {Probes of the small $x$ gluon via exclusive $J/\psi$ and $\Upsilon$
  production at HERA and the LHC}.
\newblock {\em JHEP}, 11:085, 2013.

\bibitem{Budnev:1975poe}
V.~M. Budnev, I.~F. Ginzburg, G.~V. Meledin, and V.~G. Serbo.
\newblock {The Two photon particle production mechanism. Physical problems.
  Applications. Equivalent photon approximation}.
\newblock {\em Phys. Rept.}, 15:181--281, 1975.

\end{thebibliography}
\endthebibliography

\end{document}